\documentclass[a4paper,twoside]{article}

\usepackage{epsfig}
\usepackage{subfigure}
\usepackage{calc}
\usepackage{amssymb}
\usepackage{amstext}
\usepackage{amsmath}
\usepackage{multicol}
\usepackage{pslatex}
\usepackage{apalike}
\usepackage{fancyhdr}
\usepackage{CISB}
\usepackage{url}
\usepackage[small]{caption}

\begin{document}
\onecolumn
\maketitle 
\normalsize 

\section{\uppercase{Introduction}}
\label{sec:introduction}
\noindent 
In this note we discuss the differential occurrence of intrinsically disordered proteins (IDPs) among disease-related proteins in the human proteome. On the basis of straightforward observations we raise some critical remarks about the often claimed strict association between the {\it unfoldome} \cite{Uversky2010} and the human {\it diseasome} \cite{Goh2007}. 

In the last decade a growing number of scientific publications have been devoted to IDPs; these are proteins lacking a well-defined three-dimensional tertiary structure in all or part of their polypeptide chain and existing as an ensemble of flexible conformations. It has been reported that IDPs fulfil important biological functions in the cell being involved in targeting, signalling and regulation of the cell cycle \cite{Wright1999,Dunker2001,Dunker2008,Tompa2010,Uversky2013,vanDerLee2014}. The absence of a tertiary structure suggests that these proteins do not use a lock-and-key mechanism to interact with their substrates, and a number of different mechanisms have been proposed, involving unfold-to-fold transitions and/or high specificity/low affinity interactions with target molecules \cite{Dunker2001,Uversky2002,Tompa2010}.

A growing interest is due also to the observation that IDPs might be essentially implied in cellular processes related to the development of human diseases. In particular, in a quite cited paper by Iakoucheva et al. of 2002, it was reported that cancer-related proteins are enriched in disorder. In another often quoted paper, it is reported that the fraction of IDPs is larger among proteins related to cardiovascular diseases (CVD) than in the SwissProt database, which, notably, contains proteins from all kingdoms of life \cite{Cheng2006}. Uversky et al. in 2009 compared the percentage of IDPs in several sets of proteins associated with different diseases and in a set of selected structured proteins from the Protein Data Bank \cite{Berman2000} taken as a reference. The conclusion that the fraction of IDPs in the set of structured proteins is significantly lower than in the sets associated with diseases \cite{Uversky2009}. Afterwards, many papers have linked IDPs with diseases \cite{Midic2009,Uversky2009,Uversky2014} and the $D^2$ concept ({\it disorder in disease}) has been introduced by Uversky in 2008 \cite{Uversky2008}. It has to be said that many of the above mentioned papers adopt criteria to define a protein as IDP that overestimate the absolute number of IDPs in a data set, as we have previously shown \cite{Deiana2013}.

The strict association of protein disorder with complex pathologies, based on the claim of a systematic abundance of IDPs in disease-related proteins, has reverberated, since 2002, in a growing corpus of publications. From the whole of these studies one can be tempted to conclude that disorder plays indeed an essential role in the development of diseases with complex etiology, e.g. cancer, diabetes, cardiovascular and neurodegenerative syndromes.  Looking back at the original studies and in most of the subsequent papers, the relatively high occurrence of IDPs in human disease-related proteins was compared with the occurrence of IDPs in sets of proteins from different organisms, which is an unfair reference. Therefore, it is incorrect to infer that, selecting human proteins annotated as disease-related, that corresponds to enriching in disordered proteins. As we show below, intrinsically disordered proteins have the same occurrence and distribution both among disease-related proteins and among the rest of the human proteome (HP) (figure 1). In a nutshell: there is no enrichment. 

In this note, after a brief discussion of criteria to recognise a protein as disordered, we compare the percentage of IDPs occurring in human proteins annotated as disease-related and in the rest of HP. We show that the fraction of IDPs in the two sub-sets is similar and equally distributed over classes of disorder. Indeed, HP contains more IDPs than other organisms, but we did not found a specific enrichment among the disease-related proteins. An observation that could have been done earlier and that urge for a critical attitude in reading the copious literature about the connection between unfoldome and diseasome.

\section{\uppercase{Methods}}
\label{sec:Methods}
\noindent The proteome of Homo Sapiens was downloaded from the SwissProt database, release of January 2011. According to the annotations in the database, the 20230 sequences in the HP were partitioned into two major subgroups: {\it disease-related} proteins (9110) and {\it the rest} (11120). Disease-related proteins were grouped then into subclasses, referring to four sets of diseases with a complex etiopathology, cancer (3427), cardiovascular diseases (CVD) (6645), diabetes (250) and neurodegenerative diseases (1380). Note that there are disease-related proteins that belong to more than one subclass. These subclasses were identified using the same set of keywords used by Uversky et al. in 2009. 

Disordered residues in protein sequences were identified by DISOPRED2 \cite{Ward2004}, a well-known disorder predictor that has both good {\it sensitivity} and good {\it specificity}. Moreover, in a previous study, we found that it has also a good {\it selectivity}, an often overlooked index of performance that essentially controls the number of false positives \cite{Deiana2013}. 

Usually protein sequences have been classified as intrinsically disordered if they contain at least one disordered segment longer than 30 amino acids \cite{Oldfield2005,LeGall2007}. 
This criterion does not take into account the length of the protein and can induce an overestimation of the number of disordered proteins in a given set. Consider, for example, the case of a protein made by 1000 residues, with only one disordered segment of 30 residues. This protein would be classified as disordered notwithstanding the fact it could well contain structured regions in the rest of the sequence. A more sensible criterion is that of requiring, for a protein to be an IDP, to have at least 30\% of disordered residues \cite{Gsponer2008,Deiana2013}. 

In this study we adopt the intersection of both criteria and we classify a protein as an IDP if it has at least one disordered domain longer than 30 residues and at least 30\% of residues that are predicted as disordered. 

Uncertainties in the percentages  of IDPs over classes of disorder (Figures 1 and 2) were estimated through bootstrap resampling and reported as standard errors \cite{Efron1993}.

\section{\uppercase{Results}}
\label{sec:Results}
\noindent Following the two combined criteria mentioned above, we assessed the percentage of IDPs among disease-related and the rest of proteins in the human proteome. In table 1 the estimates based on the first and the second criterion, used either separatedly or in combination, are compared. 

\begin{table}[h]
\caption{{\bf Estimates of the percentage of IDPs in disease-related and the rest of the human proteome (HP).} \\  The first or the second criterion and the combination of the two (see text) are here considered.In the first criterion a protein is considered as intrinsically disordered if it has at least one disordered segment longer than 30 consecutive amino acids; in the second criterion a protein is disordered if more than 30\% of its amino acids are predicted as disordered. Percentages are computed by normalising with respect the number of proteins in each group.}\label{tab:table1} \centering
\begin{tabular}{|c|c|c|c|}
  \hline
  GROUP & 1st Crit. & 2nd Crit. & Both criteria\\
  \hline
  & IDPs, \% & IDPs, \% & IDPs, \% \\
  \hline
  Disease-related&0.66 &0.43 & 0.41\\
  \hline
  Rest of HP& 0.63& 0.49&0.46 \\
  \hline
\end{tabular}
\vspace{-0.3cm}
\end{table}

\vfill

Clearly, the second criterion is more conservative. Interestingly, if one adopts both criteria simultaneously one gets: 41\% for the group of disease-related and 46\% for the rest of proteins in the human proteome; these figures indicate that the second criterion is essential to reject, from both groups, spurious cases of proteins that are classified as disordered just because they comprise a disordered segment longer than 30 residues, but nevertheless are not disordered in the major part of their sequences. The second criterion, that takes into account the fraction of disordered residues, reduces the incidence of IDPs in both groups by around 20\%. In the following we adopt both criteria in combination. 

It is then interesting to estimate in disease-related human proteins and the rest of HP the probability that an IDP is disordered in at least a given percentage of its residues (see Figure 1), a kind of cumulative distribution. The distribution of IDPs over bins of percentages of disorder follow similar decreasing trends in both groups. Figure 1 clearly shows that, not only the incidence of IDPs in disease-related proteins is lower than in the rest of the human proteome globally, but also that it is lower in each bin of disorder, in detail.

\vfill

\begin{figure}[!h]
  \vspace{-0.2cm}
  \centering
   {\epsfig{file = 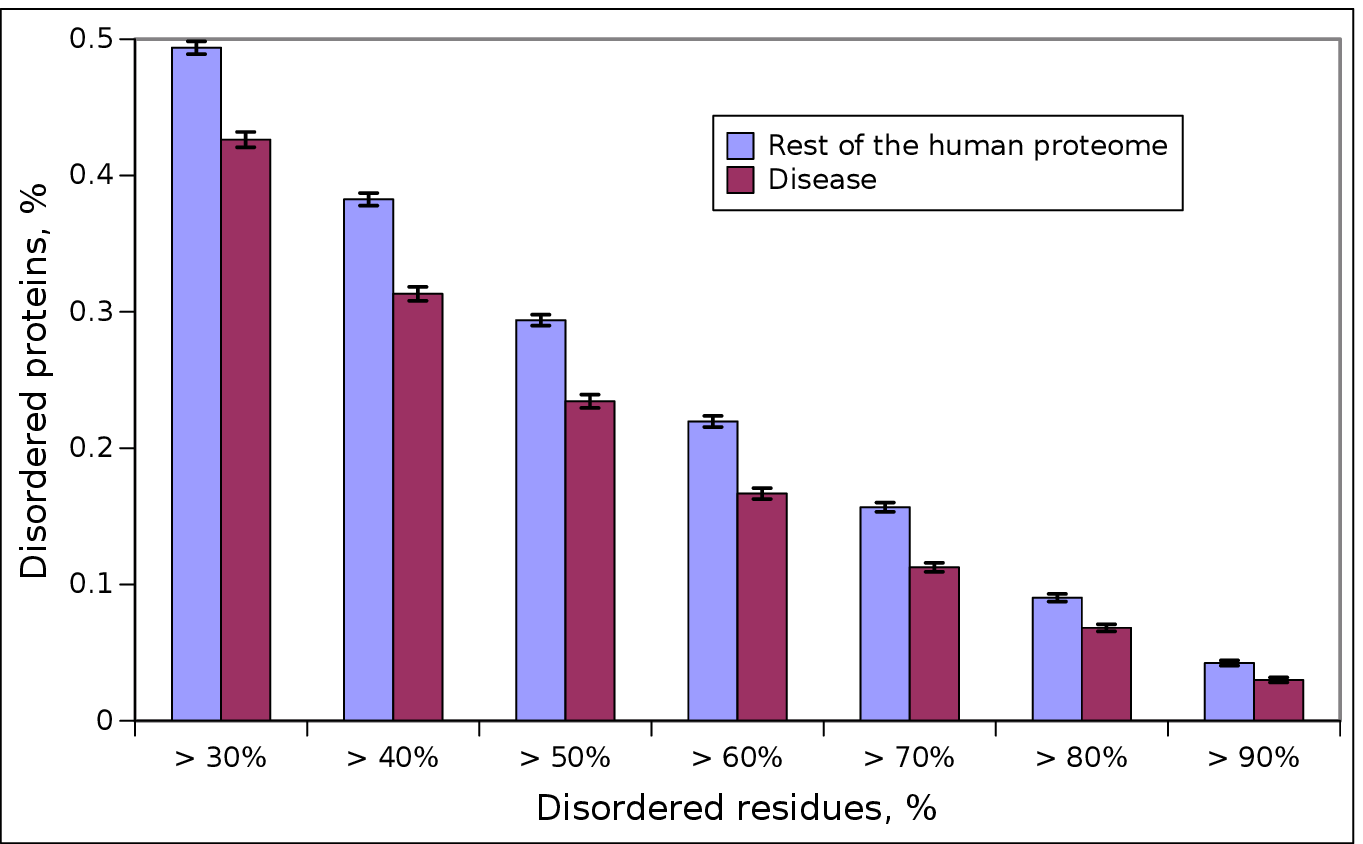, width = 10.5cm}}
  \caption{{\bf Distribution of IDPs in disease-related proteins and in the rest of the human proteome.}\\
\noindent Disorder is here binned. The first bin groups IDPs with more than 30\% of disordered residues; the second bin corresponds to proteins with more than 40\% of disordered residues and so on. Clearly the second bin contains all the proteins in the first bin minus those proteins that have between 30\% and 40\% of predicted disordered residues. The frequency of occurrences is the number of IDPs in each bin divided, respectively, by the number of disease-related proteins and by the number of remaining proteins in the HP. The error bars represent standard errors (see section Methods) upon bootstrap resampling.
}
  \label{fig1}
  \vspace{-0.5cm}
\end{figure}

\vfill

The human proteome has a large amount of IDPs, from the estimates in this note, it is close to 44\%, but we cannot say that IDPs are more frequent among disease-related proteins. Disease-related proteins cannot be thought of as a group of proteins that are specifically enriched in disorder; they are disordered as much as the rest of the human proteome. 
Other interesting observations come from considering the cumulative distribution of disease-related IDPs over classes of disorder, separating various groups of diseases (see figure 2, where the occurrence of IDPs in the rest of HP is kept as a reference).  The incidence of IDPs is different for different pathologies, though it tends to become uniform and comparable with that of the rest of HP in the classes with a large majority of disordered residues. Interestingly, in proteins related to cancer the occurrence of IDPs is higher than in the rest of the human proteome, in most classes of disorder.  It becomes comparable with the rest of HP only from the bin of 70\% on. IDPs related to neurodegenerative diseases follow next, but are less frequent than in cancer and in the rest of the HP, again becoming comparable in incidence only in the classes that are disordered for more than 80\%. Proteins related to CVDs and diabetes appear to have less incidence of IDPs with  respect to the other two subgroups and the rest of HP.

\vfill

\begin{figure}[!h]
  \vspace{-0.2cm}
  \centering
   {\epsfig{file = 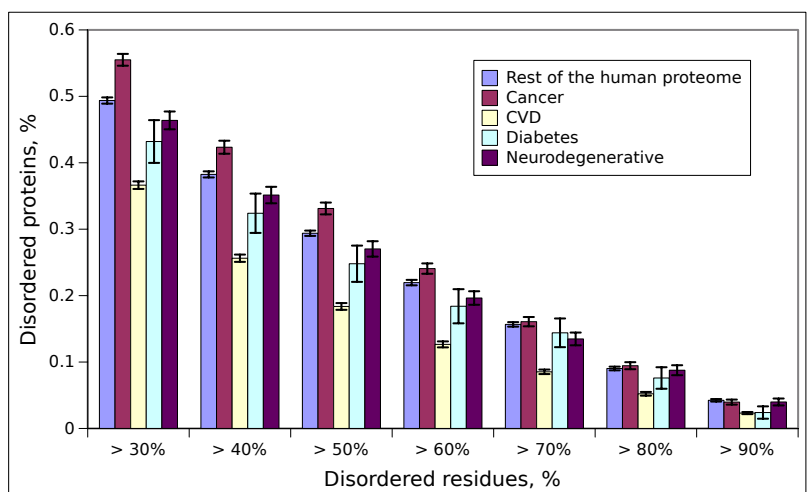, width = 10.5cm}}
  \caption{{\bf Percent distribution of disease-related IDPs in classes of disorder for different groups of diseases}.\\
  \noindent Percentages of IDPs related to different groups of diseases are here binned over the same classes of disorder as in figure 1. The occurrence of IDPs in the rest of the human proteoome is reported as a reference.}
  \label{fig2}
  \vspace{-0.5cm}
\end{figure}

\vfill

\section{\uppercase{Discussion}}
\label{sec:Discussion}
\noindent Consistently with previous reports, we confirm that the fraction of IDPs is higher in cancer-related proteins with respect to the other groups \cite{Iakoucheva2002,Uversky2009}. Nevertheless it has to be noted that in those papers the comparison between the occurrence of IDPs in cancer-related human proteins is made with: signalling proteins, proteins in Swissprot from Eukaryots, and a non redundant non-homologous set of proteins from the PDB. No comparison is made, as we do here, with the rest of the HP. The remarkable enrichment of disorder in disease-related proteins, as shown in figure 1 of the paper by Iakoucheva et al. of 2002 and in figure 4 of the 2009 paper by Uversky et al., is due to the comparison made between disease-related human proteins and sets of proteins that mix up human proteins with proteins from other organisms that, as is well known, are biased towards less disorder. Moreover, in those original works the first criterion of IDPs classification was adopted, inducing an overestimate of the amount of disorder in all the considered instances, as we have discussed above.
Notably, the biased estimates in \cite{Iakoucheva2002}, and \cite{Uversky2009} propagated and somehow flawed many papers. Also recently we have found the same way of arguing, same type of graphs, in two interesting papers on the network of interactions governing cell death processes \cite{Uversky2013,Peng2013}. 

As is well known one of the major ÒplayersÓ in the game of enforcing or escaping programmed cell death is the protein p53, which is inactivated in most cancers and has unstructured N- and C-terminal regions and a structured DNA-binding region. This protein is central to many discussions about the essential role of disorder in cancer, see, e.g., the articulate recent review by Uversky \cite{Uversky2014}. But it is worth noting that in this paper and in many that we have checked in the same thread no mention is made of the important observations  published by Pajkos et al. in 2012. In that work \cite{Pajkos2012} it is clearly suggested that there could be a problem of overestimation of disorder in previous works \cite{Iakoucheva2002,Uversky2009}. Moreover, Pajkos et al. observe, that in p53 cancer-related mutations and polymorphisms essentially occur in the structured regions and not in the disordered regions. This observation deserves due consideration in the current investigation of the role of disorder in the p53 machinery, but seems to have been overlooked.

\section{\uppercase{Conclusions}}
\label{sec:Conclusions}
\noindent Intrinsically disordered proteins undoubtedly hold a peculiar fascination for the community of protein science. Under the enthusiast pressure of a group of devoted researchers the number of publications on the theme of IDPs is growing at a remarkable pace. Nevertheless, it must be noted that while on the one hand the number of papers on IDPs that are based on predictors is growing exponentially, on the other hand the number of circumstantiated experiments on single cases is growing perhaps logarithmically in time. Since doing experiments, particularly in vivo, is more demanding than doing computations this last observation is almost trivial, one could say, but perhaps is worth noting. 

It seems that the notion of intrinsic disorder in proteins still escapes a unitary physical-chemical definition, possibly there are many variants of disorder and, nowadays after more than a decade of enthusiasm it would be necessary to make things clear. Even the term ÒdisorderÓ might be out of place and should deserve clarification and exact definitions. We observe that a well-established line of research has a certain tendency at increasing the semantics in the field, as if IDPs were playing the role of the "mysterious" dark matter of contemporary protein science \cite{Uversky2010}. From an operational point of view the concept of IDPs that is implemented in this line of research is, essentially, that of protein sequences enriched in disorder-promoting residues (i.e. charged and hydrophilic). 
Looking at the future it must be said that very recently there have been two important meetings: see: \url{http://www.grc.org/programs.aspx?year=2014&program=idp} and \url{http://www.biophysics.org/2014dublin/Home/tabid/4526/Default.aspx}) that point toward a more mature season of studies aimed at a detailed and specific definition of protein disorder (by the way, from a physical point of view even a globular crystallisable protein is a disordered amorphous molecular system, a small piece of glassy material, lacking. e.g., any periodicity).

Our task in this note was just to warn that in dealing with protein disorder one has to face a deluge of publications that should be critically considered and not enthusiastically accepted, without checking, as increasing evidence that disorder plays essential roles in fundamental biological processes. We believe that, at the moment, more sober and circumstantiate approaches have to be implemented (e.g. \cite{Babu2012}).

We conclude by saying that prevalence of disorder among disease-related proteins is not a rule. Indeed, cancer related proteins are, in our check, as statistically more disordered than the rest of the HP. In our opinion, there is no need anymore to argue that protein disorder is important because of its relative high incidence among the proteins that are in many ways related to high-impact diseases like cancer, but simply because it is an interesting and pervasive mode of protein function, particularly in HP. Disorder is common in the human proteome, but seems not a specific signature of diseases. 

\section{\uppercase{Acknowledgements}}
\noindent One of us (A.D.) has been enrolled, from 2007 to 2010, in cycle XXIII of the graduate school of Biophysics of Sapienza University of Rome, at the time part of the CISB. Both of us warmly tank prof. Alfredo Colosimo for the stimulating atmosphere and the support during those years; the study of intrinsic disorder in proteins has been the main theme of  Antonio Deiana's doctoral thesis.

\renewcommand{\baselinestretch}{0.98}
\bibliographystyle{apalike}
{\small
\bibliography{Draft_ag.bib}}
\renewcommand{\baselinestretch}{1}

\end{document}